\documentclass[conference,a4paper]{IEEEtran}

\usepackage[english]{babel}
\usepackage[utf8]{inputenc}
\usepackage[T1]{fontenc}
\usepackage{hyphenat}

\usepackage{times}
\usepackage{amsmath,amssymb,amstext,amsfonts,mathrsfs}
\usepackage{comment}

\usepackage{graphicx}                           
\graphicspath{{.}{./Figures/}}
\usepackage[tight]{subfigure}                   

\usepackage[svgnames]{xcolor}                   
\usepackage{tikz}                               
\usepackage{pgf}

\usepackage[nolist,printonlyused]{acronym}      

\usepackage[sort]{cite}                         

\usepackage{enumerate}                          

\usepackage{multirow}
\usepackage{tabularx}
\newcolumntype{C}{>{\centering\arraybackslash}X}
\newcolumntype{R}{>{\flushright\arraybackslash}X}
\newcolumntype{L}{>{\flushleft\arraybackslash}X}

\usepackage{psfrag}
\usepackage{algorithmic}
\usepackage[ruled]{algorithm}
\algsetup{indent=2ex}

\usepackage{setspace}
\usepackage{algorithmic}
\usepackage[ruled]{algorithm}
\algsetup{indent=2ex}

\usepackage{xspace} 
\newcommand*{\unit}[2]{\mbox{\ensuremath{#1\,\mathrm{#2}}}}

\usepackage{float}
\floatname{algorithm}{\footnotesize Algorithm}
\usepackage{multirow}
\usepackage{url}

\newcommand{\maxi}{\mathop{\displaystyle \mbox{maximize}}}

\newcommand*{\FigRef}[1]{Figure~\ref{#1}}
\newcommand*{\SecRef}[1]{Section~\ref{#1}}
\newcommand*{\TabRef}[1]{Table~\ref{#1}}

\newcommand*{\AlgRef}[1]{Algorithm~\ref{#1}}

\IEEEoverridecommandlockouts

\begin{document}
\date{\today}

\title{Performance Analysis of Network-Assisted Two-Hop D2D Communications}
\singlespacing

\author{
Jos\'{e} Mairton B.~da Silva Jr$^*$, G\'{a}bor Fodor$^\dagger$ Tarcisio F.~Maciel$^*$\\
$^*$Wireless Telecommunications Research Group (GTEL), Federal University of Ceará (UFC), Fortaleza, Ceará, Brazil. \\
$^\dagger$ Ericsson Research, Stockholm, Sweden. 
}


\begin{acronym}[MU-MIMO]
  \acro{2G}{Second Generation}
  \acro{3G}{3$^\text{rd}$~Generation}
  \acro{3GPP}{3$^\text{rd}$~Generation Partnership Project}
  \acro{4G}{4$^\text{th}$~Generation}
  \acro{5G}{5$^\text{th}$~Generation}
  \acro{AA}{Antenna Array}
  \acro{AC}{Admission Control}
  \acro{AD}{Attack-Decay}
  \acro{ADSL}{Asymmetric Digital Subscriber Line}
  \acro{ARMA}{Autoregressive Moving Average}
  \acro{APA}{Adaptive Power Allocation}
  \acro{ATES}{Adaptive Throughput-based Efficiency-Satisfaction Trade-Off}
  \acro{AWGN}{Additive White Gaussian Noise}
  \acro{BB}{Branch and Bound}
  \acro{BD}{Block Diagonalization}
  \acro{BER}{Bit Error Rate}
  \acro{BF}{Best Fit}
  \acro{BLER}{BLock Error Rate}
  \acro{BPSK}{Binary Phase-Shift Keying}
  \acro{BS}{Base Station}
  \acro{CAPEX}{Capital Expenditure}
  \acro{CBF}{Coordinated Beamforming}
  \acro{CBS}{Class Based Scheduling}
  \acro{CC}{Congestion Control}
  \acro{CDF}{Cumulative Distribution Function}
  \acro{CDMA}{Code-Division Multiple Access}
  \acro{CNR}{Channel-to-Noise Ratio}
  \acro{CPICH}{Common Pilot Channel}
  \acro{CoMP}{Coordinated Multi-Point}
  \acro{CQI}{Channel Quality Indicator}
  \acro{CRM}{Constrained Rate Maximization}
  \acro{CS}{Coordinated Scheduling}
  \acro{CSI}{Channel State Information}
  \acro{D2D}{Device-to-Device}
  \acro{DCA}{Dynamic Channel Allocation}
  \acro{DE}{Differential Evolution}
  \acro{DFT}{Discrete Fourier Transform}
  \acro{DIST}{Distance-based Grouping}
  \acro{DL}{Downlink}
  \acro{DMA}{Double Moving Average}
  \acro{DMS}{D2D Mode Selection}
  \acro{DPC}{Dirty Paper Coding}
  \acro{DRA}{Dynamic Resource Assignment}
  \acro{DSM}{Delay-based Satisfaction Maximization}
  \acro{ECC}{Electronic Communications Committee}
  \acro{EFLC}{Error Feedback Based Load Control}
  \acro{EI}{Efficiency Indicator}
  \acro{eNB}{Evolved Node B}
  \acro{EPA}{Equal Power Allocation}
  \acro{EPC}{Evolved Packet Core}
  \acro{EPS}{Evolved Packet System}
  \acro{E-UTRAN}{Evolved Universal Terrestrial Radio Access Network}
  \acro{FDD}{Frequency Division Duplex}
  \acro{FDM}{Frequency Division Multiplexing}
  \acro{FER}{Frame Erasure Rate}
  \acro{FSB}{Fixed Switched Beamforming}
  \acro{FTP}{File Transfer Protocol}
  \acro{GA}{Genetic Algorithm}
  \acro{GBR}{Guaranteed Bit Rate}
  \acro{GLR}{Gain to Leakage Ratio}
  \acro{GPL}{GNU General Public License}
  \acro{HARQ}{Hybrid Automatic Repeat Request}
  \acro{HMS}{Harmonic Mode Selection}
  \acro{HOL}{Head Of Line}
  \acro{HSDPA}{High-Speed Downlink Packet Access}
  \acro{HSPA}{High Speed Packet Access}
  \acro{HTTP}{HyperText Transfer Protocol}
  \acro{ICMP}{Internet Control Message Protocol}
  \acro{ICI}{Intercell Interference}
  \acro{ID}{Identification}
  \acro{IETF}{Internet Engineering Task Force}
  \acro{UID}{Unique Identification}
  \acro{IID}{Independent and Identically Distributed}
  \acro{IIR}{Infinite Impulse Response}
  \acro{ILP}{Integer Linear Problem}
  \acro{IMT}{International Mobile Telecommunications}
  \acro{INV}{Inverted Norm-based Grouping}
  \acro{IoT}{Internet of Things}
  \acro{IP}{Internet Protocol}
  \acro{IPv6}{Internet Protocol Version 6}
  \acro{ISD}{Inter-Site Distance}
  \acro{ISI}{Inter Symbol Interference}
  \acro{ITU}{International Telecommunication Union}
  \acro{JOAS}{Joint Opportunistic Assingnment and Scheduling}
  \acro{JP}{Joint Processing}
  \acro{KKT}{Karush-Kuhn-Tucker}
  \acro{LAC}{Link Admission Control}
  \acro{LA}{Link Adaptation}
  \acro{LC}{Load Control}
  \acro{LTE}{Long Term Evolution}
  \acro{LTE-Advanced}{Long Term Evolution Advanced}
  \acro{MAC}{Medium Access Control}
  \acro{MANET}{Mobile Ad-hoc Networks}
  \acro{MCS}{Modulation and Coding Scheme}
  \acro{MDB}{Measured Delay Based}
  \acro{MF}{Matched Filter}
  \acro{MG}{Maximum Gain}
  \acro{MH}{multi-hop}
  \acro{MIMO}{Multiple Input Multiple Output}
  \acro{MISO}{Multiple Input Single Output}
  \acro{MLWDF}{Modified Largest Weighted Delay First}
  \acro{MME}{Mobility Management Entity}
  \acro{MMSE}{Minimum Mean Square Error}
  \acro{MOS}{Mean Opinion Score}
  \acro{MPF}{Multicarrier Proportional Fair}
  \acro{MRA}{Maximum Rate Allocation}
  \acro{MR}{Maximum Rate}
  \acro{MRC}{Maximum Ratio Combining}
  \acro{MRT}{Maximum Ratio Transmission}
  \acro{MRUS}{Maximum Rate with User Satisfaction}
  \acro{MSE}{Mean Squared Error}
  \acro{MSI}{Multi-Stream Interference}
  \acro{MTC}{Machine-Type Communication}
  \acro{MTSI}{Multimedia Telephony Services over IMS}
  \acro{MU-MIMO}{Multi-User Multiple Input Multiple Output}
  \acro{MU}{Multi-User}
  \acro{NAS}{Non-Access Stratum}
  \acro{NB}{Node B}
  \acro{NLOS}{Non-Line of Sight}
  \acro{NSPS}{National Security and Public Safety}
  \acro{NORM}{Normalized Projection-based Grouping}
  \acro{NP}{Non-Polynomial Time}
  \acro{NRT}{Non-Real Time}
  \acro{OFDMA}{Orthogonal Frequency Division Multiple Access}
  \acro{OFDM}{Orthogonal Frequency Division Multiplexing}
  \acro{OPEX}{Operational Expenditure}
  \acro{ORB}{Orthogonal Random Beamforming}
  \acro{JO-PF}{Joint Opportunistic Proportional Fair}
  \acro{OSI}{Open Systems Interconnection}
  \acro{PAIR}{D2D Pair Gain-based Grouping}
  \acro{PAPR}{Peak-to-Average Power Ratio}
  \acro{P2P}{Peer-to-Peer}
  \acro{PC}{power control}
  \acro{PER}{Packet Error Rate}
  \acro{PF}{Proportional Fair}
  \acro{P-GW}{Packet Data Network Gateway}
  \acro{PRB}{Physical Resource Block}
  \acro{PROJ}{Projection-based Grouping}
  \acro{ProSe}{Proximity Services}
  \acro{PS}{Packet Scheduling}
  \acro{PSO}{Particle Swarm Optimization}
  \acro{PZF}{Projected Zero-Forcing}
  \acro{QAM}{Quadrature Amplitude Modulation}
  \acro{QoS}{Quality of Service}
  \acro{QPSK}{Quadri-Phase Shift Keying}
  \acro{RAISES}{Reallocation-based Assignment for Improved Spectral Efficiency and Satisfaction}
  \acro{RAN}{Radio Access Network}
  \acro{RA}{Resource Allocation}
  \acro{RATE}{Rate-based}
  \acro{RB}{Resource Block}
  \acro{RBG}{Resource Block Group}
  \acro{RLC}{Radio Link Control}
  \acro{RM}{Rate Maximization}
  \acro{RNC}{Radio Network Controller}
  \acro{RND}{Random Grouping}
  \acro{RRA}{Radio Resource Allocation}
  \acro{RRM}{Radio Resource Management}
  \acro{RSCP}{Received Signal Code Power}
  \acro{RSRP}{Reference Signal Receive Power}
  \acro{RSRQ}{Reference Signal Receive Quality}
  \acro{RR}{Round Robin}
  \acro{RRC}{Radio Resource Control}
  \acro{RSSI}{Received Signal Strength Indicator}
  \acro{RT}{Real Time}
  \acro{RU}{Resource Unit}
  \acro{RV}{Random Variable}
  \acro{SAC}{Session Admission Control}
  \acro{SCM}{Spatial Channel Model}
  \acro{SC-FDMA}{Single Carrier - Frequency Division Multiple Access}
  \acro{SD}{Soft Dropping}
  \acro{SDMA}{Space-Division Multiple Access}
  \acro{SER}{Symbol Error Rate}
  \acro{SES}{Simple Exponential Smoothing}
  \acro{S-GW}{Serving Gateway}
  \acro{SINR}{Signal to Interference-plus-Noise Ratio}
  \acro{SI}{Satisfaction Indicator}
  \acro{SIP}{Session Initiation Protocol}
  \acro{SISO}{Single Input Single Output}
  \acro{SIMO}{Single Input Multiple Output}
  \acro{SIR}{Signal to Interference Ratio}
  \acro{SMA}{Simple Moving Average}
  \acro{SNR}{Signal to Noise Ratio}
  \acro{SORA}{Satisfaction Oriented Resource Allocation}
  \acro{SORA-NRT}{Satisfaction-Oriented Resource Allocation for Non-Real Time Services}
  \acro{SORA-RT}{Satisfaction-Oriented Resource Allocation for Real Time Services}
  \acro{SPF}{Single-Carrier Proportional Fair}
  \acro{SRA}{Sequential Removal Algorithm}
  \acro{SU-MIMO}{Single-User Multiple Input Multiple Output}
  \acro{SU}{Single-User}
  \acro{SVD}{Singular Value Decomposition}
  \acro{TCP}{Transmission Control Protocol}
  \acro{TDD}{Time Division Duplex}
  \acro{TDMA}{Time Division Multiple Access}
  \acro{ThMS}{Threshold-based Mode Selection}
  \acro{TTI}{Transmission Time Interval}
  \acro{TSM}{Throughput-based Satisfaction Maximization}
  \acro{TU}{Typical Urban}
  \acro{UE}{User Equipment}
  \acro{UEPS}{Urgency and Efficiency-based Packet Scheduling}
  \acro{UL}{Uplink}
  \acro{UMTS}{Universal Mobile Telecommunications System}
  \acro{URI}{Uniform Resource Identifier}
  \acro{URM}{Unconstrained Rate Maximization}
  \acro{VR}{Virtual Resource}
  \acro{VoIP}{Voice over IP}
  \acro{WCDMA}{Wideband Code Division Multiple Access}
  \acro{WF}{Water-filling}
  \acro{WiMAX}{Worldwide Interoperability for Microwave Access}
  \acro{WINNER}{Wireless World Initiative New Radio}
  \acro{WMPF}{Weighted Multicarrier Proportional Fair}
  \acro{WPF}{Weighted Proportional Fair}
  \acro{WWW}{World Wide Web}
  \acro{XIXO}{(Single or Multiple) Input (Single or Multiple) Output}
  \acro{ZF}{Zero-Forcing}
  \acro{ZMCSCG}{Zero Mean Circularly Symmetric Complex Gaussian}
  \acro{2G}{Second Generation}
  \acro{3GPP}{3$^\text{rd}$~Generation Partnership Project}
  \acro{3G}{Third Generation}
  \acro{4G}{4$^\text{th}$~Generation}
  \acro{5G}{5$^\text{th}$~Generation}
  \acro{4G-WEST}{Fourth Generation - Wireless Evolution Simulation Tool}
  \acro{16QAM}{16 Quadrature Amplitude Modulation}
  \acro{64QAM}{64 Quadrature Amplitude Modulation}
  \acro{AA}{Antenna Array}
  \acro{ABC}{Always Best Connected}
  \acro{AC}{Admission Control}
  \acro{ACG}{Amplitude Craving Greedy}
  \acro{ACK}{Acknowledgement}
  \acro{ACS}{Access Control Server}
  \acro{AcVI}{Actual Value Interface}
  \acro{A-DPCH}{Associated Dedicated Physical Channel}
  \acro{ADS}{Asymptotic Delay Scheduler}
  \acro{ADSL}{Asymmetric Digital Subscriber Line}
  \acro{AF}{Amplify-and-Forward}
  \acro{AG}{Absolute Grant}
  \acro{aGW}{Access Gateway}
  \acro{AICH}{Acquisition Indication Channel}
  \acro{AM}{Acknowledged Mode}
  \acro{AMC}{Adaptive Modulation and Coding}
  \acro{AMR}{Adaptive Multirate}
  \acro{APA}{Adaptive Power Allocation}
  \acro{ARQ}{Automatic Repeat Request}
  \acro{AS}{Active Set}
  \acro{AVI}{Average Value Interface}
  \acro{AWGN}{Additive White Gaussian Noise}
  \acro{B3G}{Beyond Third Generation}
  \acro{BABS}{Bandwidth Assignment Based on SNR}
  \acro{BB}{Branch-and-Bound}
  \acro{BCH}{Broadcast Channel}
  \acro{BD}{Block Diagonalization}
  \acro{BER}{Bit Error Rate}
  \acro{BF}{Best Fit}
  \acro{BI-OFDMA}{Block Interleaved Orthogonal Frequency Division Multiple Access}
  \acro{BLDPCC}{Block Low Density Parity Check Codes}
  \acro{BLEP}{Block Error Probability}
  \acro{BLER}{Block Error Rate}
  \acro{B-OFDMA}{Block Orthogonal Frequency Division Multiple Access}
  \acro{BPSK}{Binary Phase-Shift Keying}
  \acro{BR}{Bandwidth Request}
  \acro{BRA}{Balanced Random Allocation}
  \acro{BS}{Base Station}
  \acro{BSA}{Best Subcarrier Allocation}
  \acro{BTC}{Block Turbo Code}
  \acro{CAC}{Call Admission Control}
  \acro{CAPEX}{Capital Expenditure}
  \acro{CC}{Chase Combining}
  \acro{CCI}{Co-Channel Interference}
  \acro{CDC}{Common Data Channel}
  \acro{CDF}{Cumulative Distribution Function}
  \acro{CDMA}{Code-Division Multiple Access}
  \acro{CDMA2000 EV-DO}{Code-Division Multiple Access 2000 Evolution-Data only}
  \acro{CESM}{Capacity ESM}
  \acro{CF}{Compress-and-Forward}
  \acro{CFN}{Connection Frame Number}
  \acro{CI}{CRC Indicator}
  \acro{CID}{Connection ID}
  \acro{CINR}{Carrier to Interference plus Noise Ratio}
  \acro{CIR}{Carrier-to-Interference Ratio}
  \acro{CL}{Closed-loop}
  \acro{CM}{Code Multiplexing}
  \acro{CMC}{Connection Mobility Control}
  \acro{CN}{Core Network}
  \acro{CoC}[CC]{Convolutional Code}
  \acro{CoopRRM}{Cooperative Radio Resource Management}
  \acro{COP}{Combinatorial Optimization Problem}
  \acro{CP}{Cyclic Prefix}
  \acro{CPA}{Cellular Protection Allocation}
  \acro{CPICH}{Common Pilot Channel}
  \acro{CQE}{Channel Quality Estimation}
  \acro{CQI}{Channel Quality Indicator}
  \acro{CRA}{Capacity-driven Resource Allocation}
  \acro{CRC}{Cyclic Redundancy Check}
  \acro{CRESM}{Cutoff Rate ESM }
  \acro{CRNC}{Controlling Radio Network Controller}
  \acro{CRRM}{Common Radio Resource Management}
  \acro{CSE}{Circuit Switched Equivalent}
  \acro{CSI}{Channel State Information}
  \acro{CTC}{Convolutional Turbo Code}
  \acro{CVS}{Concurrent Versions System}
  \acro{D2D}{Device-to-Device}
  \acro{DAB}{Digital Audio Broadcasting}
  \acro{DAC}{Direct Access Channel}
  \acro{DoA}{Direction-of-Arrival}
  \acro{DBTC}{Duo-Binary Turbo Codes}
  \acro{DCA}{Dynamic Channel Allocation}
  \acro{DFT}{Discrete Fourier Transform}
  \acro{DF}{Decode-and-Forward}
  \acro{DS}{Delay Scheduler}
  \acro{DSA}{Dynamic Subcarrier Allocation}
  \acro{DCH}{Dedicated Channel}
  \acro{DC}{Direct Current}
  \acro{DL}{Downlink}
  \acro{DL-MAP}{Downlink Map}
  \acro{DL-SCH}{Downlink Shared Channel}
  \acro{DLUTRANSIM}[DL UTRANSim]{R5 Downlink UTRAN Simulator}
  \acro{DPCCH}{Dedicated Physical Control Channel}
  \acro{DPCH}{Dedicated Physical Channel}
  \acro{DRA}{Dynamic Resource Allocation}
  \acro{DRRA}{Dynamic Radio Resource Allocation}
  \acro{DTFT}{Discrete Time Fourier Transform}
  \acro{DTX}{Discontinuous Transmission}
  \acro{DVB-T}{Digital Video Broadcasting - Terrestrial}
  \acro{E-AGCH}{E-DCH Absolute Grant Channel}
  \acro{EC}{Encryption Control}
  \acro{E-DCH}{Enhanced Dedicated Channel}
  \acro{E-DPCCH}{E-DCH Dedicated Physical Control Channel}
  \acro{E-DPDCH}{E-DCH Dedicated Physical Data Channel}
  \acro{EESM}{Exponential ESM}
  \acro{E-HICH}{E-DCH Hybrid ARQ Indicator Channel}
  \acro{eNB}{Enhanced Node B}
  \acro{EPA}{Equal Power Allocation}
  \acro{EPC}{Evolved Packet Core}
  \acro{E-RGCH}{E-DCH Relative Grant Channel}
  \acro{ESM}{Effective SNR Mapping}
  \acro{E-TF}{E-DCH Transport Format}
  \acro{E-TFC}{E-DCH Transport Format Combination}
  \acro{E-TFS}{E-DCH Transport Format Set}
  \acro{ETSI}{European Telecommunications Standards Institute}
  \acro{ETU}{Extended Typical Urban}
  \acro{EUL}[E-UL]{WCDMA Enhanced Uplink}
  \acro{E-UTRA}{Evolved UMTS Terrestrial Radio Access}
  \acro{E-UTRAN}{Evolved UMTS Terrestrial Radio Access Network}
  \acro{FDD}{Frequency Division Duplex}
  \acro{FDE}{Frequency-Domain Equalization}
  \acro{FDM}{Frequency Division Multiplexing}
  \acro{FDMA}{Frequency Division Multiple Access}
  \acro{FEC}{Forward Error Correction}
  \acro{FER}{Frame Erasure Rate}
  \acro{FFT}{Fast Fourier Transform}
  \acro{FIFO}{First-In-First-Out}
  \acro{FP}{Frame Protocol}
  \acro{FRF}{Frequency Reuse Factor}
  \acro{FSB}{Fixed Switched Beams}
  \acro{FST}{Fixed SNR Target}
  \acro{GoB}{Grid of Beams}
  \acro{FSQP}{Feasible Sequential Quadratic Programming}
  \acro{FTP}{File Transfer Protocol}
  \acro{GA}{Genetic Algorithm}
  \acro{GAA}{Global Allocation Algorithm}
  \acro{GERAN}{GSM/EDGE Radio Access Network}
  \acro{GGSN}{Gateway GPRS Support Node}
  \acro{GINR}{Gain-to-Interference plus Noise Ratio}
  \acro{GSM}{Global System for Mobile communication}
  \acro{GTEL}{Wireless Telecommunications Research Group}
  \acro{HARQ}{Hybrid Automatic Repeat Request}
  \acro{HO}{Handover}
  \acro{HHO}{Hard Handover}
  \acro{HSDPA}{High Speed Downlink Packet Access}
  \acro{HS-DPCCH}{High Speed Dedicated Physical Control Channel}
  \acro{HS-DSCH}{High Speed Downlink Shared Channel}
  \acro{HS-SCCH}{High Speed Shared Control Channel}
  \acro{HSPA}{High Speed Packet Access}
  \acro{HT}{Hilly Terrain}
  \acro{HTTP}{Hypertext Transfer Protocol}
  \acro{ICI}{Intercarrier Interference}
  \acro{ICIC}{Inter-Cell Interference Coordination}
  \acro{ID}{Identifier}
  \acro{IDFT}{Inverse Discrete Fourier Transform}
  \acro{IDMA}{Interleaved Division Multiple Access}
  \acro{IEEE}{Institute of Electrical and Electronics Engineers}
  \acro{IFDMA}{Interleaved Frequency Division Multiple Access}
  \acro{IFFT}{Inverse Fast Fourier Transform}
  \acro{ILPC}{Inner-loop Power Control}
  \acro{IMS}{IP Multimedia Subsystem}
  \acro{IMT}{International Mobile Telecommunications}
  \acro{IMT-A}{International Mobile Telecommunications - Advanced}
  \acro{Ioc}{Other Cell Interference}
  \acro{IP}{Internet Protocol}
  \acro{IPR}{Intellectual Property Rights}
  \acro{IPv4}{Internet Protocol Version 4}
  \acro{IPv6}{Internet Protocol Version 6}
  \acro{IQ}{In-Phase and Quadrature}
  \acro{IR}{Incremental Redundancy}
  \acro{IRC}{Interference Rejection Combining}
  \acro{ISI}{Inter Symbol Interference}
  \acro{ISR}{Interference-to-Signal Ratio}
  \acro{IST}{Information Society Technologies}
  \acro{KKT}{Karush-Kuhn-Tucker}
  \acro{L1}{Physical Layer}
  \acro{L2}{Data Link Layer}
  \acro{L3}{Network Layer}
  \acro{L4}{Transport Layer}
  \acro{L7}{Application Layer}
  \acro{L2S}{Link-to-System}
  \acro{LA}{Link Adaptation}
  \acro{LB}{Load Balancing}
  \acro{LC}{Load Control}
  \acro{LDPCC}{Low Density Parity Check Codes}
  \acro{LDS}{Linear Delay Scheduler}
  \acro{LESM}{Logarithmic ESM}
  \acro{LiESM}{Linear ESM}
  \acro{LLR}{Log-Likelihood Ratio}
  \acro{LLSimO}{Link-Level Simulator OFDM-based}
  \acro{LuT}{Look-up Table}
  \acro{LLWCDMAEUL}[LL WCDMA E-UL]{Link-level WCDMA E-UL}
  \acro{LTE}{Long Term Evolution}
  \acro{LTE-A}{Long Term Evolution Advanced}
  \acro{MA}{Margin Adaptive}
  \acro{MAC}{Medium Access Control}
  \acro{MAP}{Maximum A Posteriori}
  \acro{MBMS}{Multimedia Broadcast Multicast Service}
  \acro{MC}{Multicarrier}
  \acro{MCH}{Multicast Channel}
  \acro{MCCH}{Multicast Control Channel}
  \acro{MCG}{Maximum Channel Gain}
  \acro{MCS}{Modulation and Coding Scheme}
  \acro{MEDS}{Method of Exact Doppler Spread}
  \acro{MG}{Maximum Gain}
  \acro{MIESM}{Mutual-Information ESM}
  \acro{MIMO}{Multiple Input Multiple Output}
  \acro{MISO}{Multiple Input Single Output}
  \acro{MME}{Mobility Management Entity}
  \acro{MPF}{Multicarrier Proportional Fair}
  \acro{MR}{Maximum Rate}
  \acro{MRC}{Maximal Ratio Combining}
  \acro{MS}{Mode Selection}
  \acro{MGINR}{Maximum GINR}
  \acro{MSINR}{Maximum SINR}
  \acro{MSS}{Maximum Segment Size}
  \acro{MURPA}{Multiuser Residual Power Allocation}
  \acro{MT}{Mobile Terminal}
  \acro{MU}{Multiuser}
  \acro{MUD}{Multiuser Diversity}
  \acro{NACK}{Negative Acknowledgement}
  \acro{NAS}{Non-Access Stratum}
  \acro{NBAP}{Node B Application Part}
  \acro{NLOS}{Non-Line-of-Sight}
  \acro{NP}{Non-deterministic Polynomial-time}
  \acro{NPC}{No Power Control}
  \acro{NR}{Noise Rise}
  \acro{NRT}{Non-Real Time}
  \acro{NSR}{Noise-to-Signal Ratio}
  \acro{NUM}{Network Utility Maximization}
  \acro{n-BSA}{n-Best Subcarrier Allocation}
  \acro{OBF}{Opportunistic Beamforming}
  \acro{OFDM}{Orthogonal Frequency Division Multiplexing}
  \acro{OFDMA}{Orthogonal Frequency Division Multiple Access}
  \acro{OFPC}{Open Loop with Fractional Path Loss Compensation}
  \acro{OL}{Open-loop}
  \acro{OLPC}{Outer-loop Power Control}
  \acro{OOP}{Object-Oriented Programming}
  \acro{OPEX}{Operational Expenditure}
  \acro{ORB}{Orthogonal Random Beamforming}
  \acro{OSA}{Opportunist Subcarrier Allocation}
  \acro{OSI}{Open Systems Interconnection}
  \acro{OSAS}{Optimal sub-channel allocation scheme}
  \acro{OVSF}{Orthogonal Variable Spreading Factor}
  \acro{PA}{Power Allocation}
  \acro{PAPR}{Peak-to-Average Power Ratio}
  \acro{P-CCPCH}{Primary Common Control Physical Channel}
  \acro{P-CPICH}{Primary Common Pilot Channel}
  \acro{PDCP}{Packet Data Convergence Protocol}
  \acro{PDCCH}{Physical Downlink Control Channel}
  \acro{PDSCH}{Physical Downlink Shared Channel}
  \acro{PDU}{Protocol Data Unit}
  \acro{PER}{Packet Error Probability}
  \acro{PF}{Proportional Fair}
  \acro{P-GW}{Packet Data Network Gateway}
  \acro{PhCH}{Physical Channel}
  \acro{PHY}{Physical}
  \acro{PICH}{Paging Indicator Channel}
  \acro{PL}{Path Loss}
  \acro{PN}{Pseudo-random Noise}
  \acro{PRB}{Physical Resource Block}
  \acro{PSC}{Packet Scheduling}
  \acro{PSTN}{Public Switched Telephone Network}
  \acro{PUSC}{Partially Used Subcarriers}
  \acro{QAM}{Quadrature Amplitude Modulation}
  \acro{QBMC}{Queue Based Max CIR}
  \acro{QBMG}{Queue-Based Maximum Gain}
  \acro{QCI}{QoS Class Identifier}
  \acro{QoS}{Quality of Service}
  \acro{QP}{Quadratic Programming}
  \acro{QPP}{Quadratic Permutation Polynomial}
  \acro{QPSK}{Quadri-Phase Shift Keying}
  \acro{R99}{Release~99}
  \acro{R5}{Release~5}
  \acro{R6}{Release~6}
  \acro{RA}{Rate Adaptive}
  \acro{RAB}{Radio Access Bearer}
  \acro{RAC}{Radio Admission Control}
  \acro{RAISES}{Reallocation-based Assignment for Improved Spectral Efficiency  and Satisfaction}
  \acro{RAN}{Radio Access Network}
  \acro{RAT}{Radio Access Technology}
  \acro{RB}{Resource Block} 
  \acro{RBC}{Radio Bearer Control}
  \acro{RBER}{Raw Bit Error Rate}
  \acro{RBIR}{Received Bit Information Rate}
  \acro{RC}{Radio Configuration}
  \acro{RET}{Remote Electronic Antenna Tilting}
  \acro{RG}{Relative Grant}
  \acro{RIP}{Radio Interface Protocol}
  \acro{RLC}{Radio Link Control}
  \acro{RLS}{Radio Link Set}
  \acro{RM}{Rate Maximization}
  \acro{RN}{Relay-based Node}
  \acro{RNA}{Radio Network Architecture}
  \acro{RNC}{Radio Network Controller}
  \acro{ROHC}{Robust Header Compression}
  \acro{RoT}{Rise over Thermal}
  \acro{RR}{Round Robin}
  \acro{RRA}{Radio Resource Allocation}
  \acro{RRC}{Radio Resource Control}
  \acro{RRM}{Radio Resource Management}
  \acro{RRV}{Rate Requirement Violation ratio}
  \acro{RSA}{Random Subcarrier Allocation}
  \acro{RS}{Resource Scheduler}
  \acro{RSN}{Retransmission Sequence Number}
  \acro{RSSI}{Received Signal Strength Indicator}
  \acro{RT}{Real Time}
  \acro{RTCP}{Real Time Control Protocol}
  \acro{RTP}{Real-Time Transport Protocol}
  \acro{RTWP}{Received Total Wideband Power}
  \acro{RU}{Resource Unit}
  \acro{RUNE}{Rudimentary Network Emulator}
  \acro{RV}{Redundancy Version}
  \acro{SA}{Simulated Annealing}
  \acro{SAW}{Stop-And-Wait}
  \acro{S-CCPCH}{Secondary Common Control Physical Channel}
  \acro{SC-FDMA}{Single Carrier - Frequency Division Multiple Access}
  \acro{SCH}{Synchronization Channel}
  \acro{SCM}{Spatial Channel Model}
  \acro{SDMA}{Space Division Multiple Access}
  \acro{SDP}{Software Development Process}
  \acro{SDU}{Service Data Unit}
  \acro{SER}{Symbol Error Rate}
  \acro{SGSN}{Service GPRS Support Node}
  \acro{S-GW}{Serving Gateway}
  \acro{SHO}{Soft Handover}
  \acro{SID}{Silence Insertion Description}
  \acro{SINR}{Signal to Interference-Plus-Noise-Ratio}
  \acro{SIP}{Session Initiation Protocol}
  \acro{SIR}{Signal-to-Interference Ratio}
  \acro{SISO}{Single Input Single Output}
  \acro{SLA}{Standard Linear Array}
  \acro{SLC}{Service Level Control}
  \acro{SM}{Subcarrier Matching}
  \acro{SNR}{Signal-to-Noise Ratio}
  \acro{SORA}{Satisfaction Oriented Resource Allocation}
  \acro{SORA-NRT}{SORA - Non Real Time}
  \acro{SORA-VoIP}{SORA - Voice over Internet Protocol}
  \acro{SORA-RT}{SORA - Real Time}
  \acro{SORA-Multi}{SORA - Multi-Service}
  \acro{SOVA}{Soft-output Viterbi Algorithm}
  \acro{SOST}{Simplified OFDMA Simulator Tool}
  \acro{SPF}{Single-Carrier Proportional Fair}
  \acro{SQP}{Sequential Quadratic Programming}
  \acro{SRF}{Sub-channel Reuse Factor}
  \acro{SRNC}{Serving Radio Network Controller}
  \acro{SS}{Subscriber Station}
  \acro{SSAS}{Simplified Subchannel Allocation Scheme}
  \acro{STBC}{Space Time Block Code}
  \acro{STFC}{Space Time Frequency Code}
  \acro{SVD}{Singular Value Decomposition}
  \acro{TB}{Transport Block}
  \acro{TC}{Turbo Code}
  \acro{TCP}{Transport Control Protocol}
  \acro{TCUU}{Traffic Channel Utilization Utility}
  \acro{TDD}{Time Division Duplexing}
  \acro{TDM}{Time Division multiplexing}
  \acro{TDMA}{Time Division Multiple Access}
  \acro{TF}{Transport Format}
  \acro{TFC}{Transport Format Combination}
  \acro{TFCS}{Transport Format Combination Set}
  \acro{TFRC}{Transport Format and Resource Combination}
  \acro{TFS}{Transport Format Set}
  \acro{TM}{Transparent Mode}
  \acro{TMU}{Throughput Marginal Utility}
  \acro{TPC}{Transmit Power Control}
  \acro{TR03}{3$^{\mathrm{rd}}$ Technical Report}
  \acro{TSN}{Transmission Sequence Number}
  \acro{TTI}{Transmission Time Interval}
  \acro{TU}{Typical Urban}
  \acro{UDP}{User Datagram Protocol}
  \acro{UE}{User Equipment}
  \acro{UEAR}{User Equipment Arrival Rate}
  \acro{UL}{Uplink}
  \acro{ULA}{Uniform Linear Array}
  \acro{ULUTRANSIM}[UL UTRANSim]{R6 Uplink UTRAN Simulator}
  \acro{UM}{Unacknowledged Mode}
  \acro{UML}{Unified Modeling Language}
  \acro{UMTS}{Universal Mobile Telecommunications System}
  \acro{UPE}{User Plane Entity}
  \acro{UTRAN}{UMTS Terrestrial Radio Access Network}
  \acro{VoIP}{Voice over IP}
  \acro{VRB}{Virtual Resource Block}
  \acro{WCDMAEUL}[WCDMA E-UL]{WCDMA Enhanced Uplink}
  \acro{WCDMA}{Wideband CDMA}
  \acro{WF}{Water Filling}
  \acro{WiMAX}{Worldwide Interoperability for Microwave Access}
  \acro{WINNER}{Wireless World Initiative New Radio}
  \acro{WLAN}{Wireless Local Area Network}
  \acro{WMPF}{Weighted Multicarrier Proportional Fair}
  \acro{WP}{Work Package}
  \acro{WP2}{Work Package Two}
  \acro{WPF}{Weighted Proportional Fair}
  \acro{WSS}{Wide-Sense Stationary}
  \acro{WWW}{World Wide Web}
  \acro{XG}{Next Generation}
  \acro{ZF}{Zero Forcing}
\end{acronym}

\singlespacing
\maketitle
\IEEEpeerreviewmaketitle

\begin{abstract}
Network-assisted single-hop device-to-device (D2D) communication can increase
the spectral and energy efficiency of cellular networks by taking advantage
of the proximity, reuse, and hop gains.
In this paper we argue that \acs{D2D} technology can be used to further increase the
spectral and energy efficiency if the key \acs{D2D} radio resource management algorithms
are suitably extended to support network assisted \textbf{\textit{multi-hop}} \acs{D2D} communications.
Specifically we propose a novel, distributed utility maximizing \acs{D2D} power control (PC)
scheme that is able to balance spectral and energy efficiency while
taking into account mode selection and resource allocation constraints that are
important in the integrated cellular-D2D environment.
Our analysis and numerical results indicate that multi-hop \acs{D2D} communications combined with the proposed
PC scheme can be useful
not only for harvesting the potential gains previously identified in the literature,
but also for extending the coverage of cellular networks.
\end{abstract}

\newcommand*{\BS}[1]{\ensuremath{\text{BS}_{#1}}}
\newcommand*{\DD}[1]{\ensuremath{\text{D2D}_\text{#1}}}
\newcommand*{\UE}[1]{\ensuremath{\text{UE}_{#1}}}

\section{Introduction}\label{sec:intro}
Although the ideas of integrating ad hoc relaying systems into cellular networks are not new
\cite{Wu-01,Fitzek-06},
the advantages of \ac{D2D} communications in cellular spectrum have been identified and
analyzed only recently \cite{Fodor2012,WenAndZhu2013}.
Specifically, it has been found that \ac{D2D} communications can increase the spectral and energy efficiency
by taking advantage of the proximity, reuse and hop gains when radio resources are properly
allocated to the cellular and \ac{D2D} layers \cite{Asadi2013}.

Another line of research suggests that relay-assisted \ac{MH} communications, including mobile relays and
relay-assisted \ac{D2D} communications can not only enhance the achievable transmission capacity, but can also improve
the coverage of cellular networks
\cite{WenAndZhu2013,Zafar-12,ChenAndHuang2013,Romero2013}.

Recognizing the potential of combining \ac{D2D} and relay technologies, the standardization and research communities
have initiated studies on the achievable gains and enabling technology components to support network-assisted
\ac{MH} \ac{D2D} communications in operator licensed spectrum.
For example, the \ac{3GPP} is investigating the use of
\ac{D2D} communication both in commercial and \ac{NSPS} scenarios \cite{3gpp.22.803}.
Integrating \ac{MH} \ac{D2D} communications can also help to meet the evolving requirements of next generation wireless
networks \cite{METIS-D1.1}.
In all these cases, both spectral and energy efficiency requirements must be met due to the limited spectrum
resources and the requirement on providing broadband services.

\begin{figure}[t!]
\vspace{8mm}
\centering
\includegraphics[width=0.90\hsize]{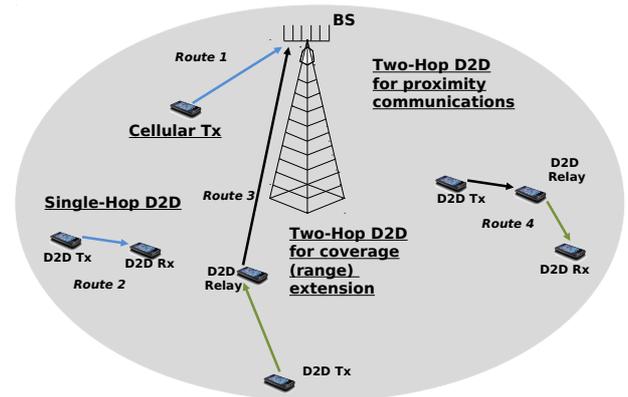}
\caption{An example of a cellular network supporting single- and multi-hop \ac{D2D} communications in cellular spectrum.
Between each source-destination (S-D) pair, a {\em route} must be defined and resources need to be allocated
to each {\em link} along the route. We use different colors to indicate different time-frequency resources,
while the same color for different links indicate the possibility for intracell resource reuse.
In this paper we assume that in the multi-hop case, the incoming and outgoing
links of a relay node must use orthogonal resources. Notice that a given S-D pair may have the possibility to
communicate in {\em cellular} mode through the base station or using single- or multi-hop \ac{D2D} communications.
}
\label{Fig:Scenario}
\end{figure}

However, extending the key enabling technology components of single-hop network-assisted \ac{D2D} communications
to \ac{MH} \ac{D2D} communication is non-trivial, because (Figure \ref{Fig:Scenario}):
\begin{enumerate}
\item Existing single-hop {\em mode selection} (MS) algorithms must be extended to
select between the single-hop \ac{D2D} link, \ac{MH} \ac{D2D} paths and cellular communications.

\item Existing single-hop {\em resource allocation} algorithms must be further developed to be able not only to manage spectrum resources between cellular and \ac{D2D} layers,
but also to comply with resource constraints along \ac{MH} paths.
\item Available \ac{D2D} \ac{PC} algorithms must be made capable of taking into account the
rate constraints of \ac{MH} paths. Specifically, it must be taken into account that along the
multiple links of a given path, only a single rate can be sustained without requiring large
buffers or facing buffer underflow situations at intermediate nodes.
\end{enumerate}

In this paper we (1) propose and analyze heuristic mode selection and resource allocation
strategies that are applicable in cellular networks integrating \ac{MH} \ac{D2D} communications and
(2) develop a utility optimal
distributed \ac{PC} scheme that takes into account both the achievable rates along
\ac{MH} paths and the overall energy consumption.
The \ac{PC} scheme can operate in concert with both the \ac{PC} schemes available in cellular networks
and the mode selection and resource allocation algorithms,
taking into account that a relaying device cannot receive and transmit data on the same
frequency resource at the same time.
Therefore, our main contribution is the \ac{MH} power control scheme that is analyzed by means
of a realistic system simulator when performing practically feasible mode selection and
resource allocation.

\section{System Model}
The system model consists of two parts. First, the {\em routing matrix}
describes the network topology and associates links with resources.
Secondly, the {\em utility function} associated with an S-D pair characterizes
the utility of supporting some communication rate between the end nodes of the pair.

\label{sec:two_hop_modelling}
\begin{figure}[!h]
\centering
\includegraphics[width=0.90\hsize]{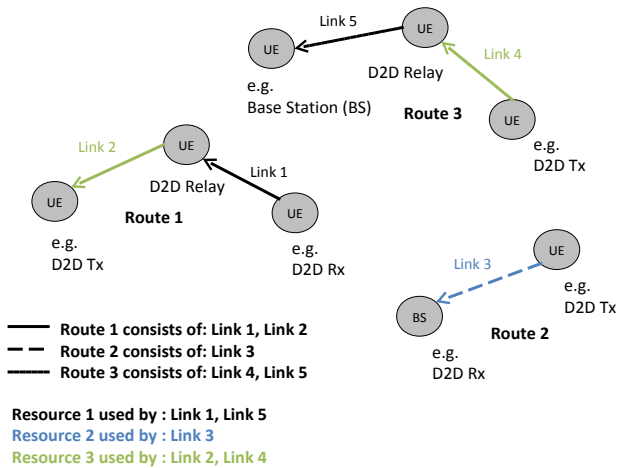}
\caption{An example of a network with 3 routes, where Route 1 and Route 3
are two-hop routes, and Route 2 consists of a single-hop route.
In the specific case of Figure \ref{Fig:Scenario},
Route 1, 2 and 3 can model the two-hop \ac{D2D} route for coverage extension,
the single-hop \ac{D2D} link and the two-hop \ac{D2D} route for proximity communication.
Note that the resources allocated to the incoming and outgoing links of a relay
node must be orthogonal, as indicated in this Figure. A node can represent a
\acl{UE} (UE) or a \acl{BS} (BS).
}
\label{Fig:Model}
\end{figure}

\subsection{Network Topology}
\label{sub:sys_model}
We model the integrated cellular-\ac{D2D} network as a set of $L$ transmitter-receiver (Tx-Rx) pairs.
A Tx-Rx pair can be a cellular \ac{UE} transmitting to its serving \ac{BS},
a \ac{D2D} Tx node transmitting to a \ac{D2D} Rx node in single-hop \ac{D2D} mode, a \ac{D2D} Tx node transmitting to
a \ac{D2D} relay node or a \ac{D2D} relay node transmitting to a \ac{D2D} Rx node.
A {\em link} refers to a single-hop transmission between a Tx-Rx pair, while a {\em route} is a concatenation
of one or more links between a S-D pair.
For example, a two-hop route consists of two Tx-Rx pairs, in which case the middle node must be a \ac{D2D}-capable relay
node (Figure \ref{Fig:Model}).
The links and routes are labelled as $l=1,\ldots,L$ and $i=1,\ldots,I$ respectively.
Next, we define the 3-dimensional {\it routing matrix} that associates links with routes and resources
and thereby describes both the network topology in terms of links and routes and the resources assigned to links.
The routing matrix is defined as
$\mathbf {R}=[r_{liq}]\in \{0,1\}^{L\times I\times Q}$, where the entry $r_{liq}$ is 1 if
data between the S-D pair $i$ is routed across link $l$ and resource $q$, and zero otherwise.
With this definition, the routing matrix can be seen as a set of $Q$ single-resource matrices,
$\mathbf {R}_q \in \{0,1\}^{L\times I}$, such that the $r_{l,i}$ element of $\mathbf{R}_q$
indicates whether link $l$ is part of route $i$ on resource $q$.
For the example of \FigRef{Fig:Model}, the $Q=3$ routing matrices are the following:
\begin{align*}
\mathbf{R}_1 &=
\begin{pmatrix}
1   & 0 & 0\cr
0   & 0 & 0\cr
0   & 0 & 0\cr
0   & 0 & 0\cr
0   & 0 & 1\cr
\end{pmatrix},\,
\mathbf{R}_2 =
\begin{pmatrix}
0   & 0 & 0\cr
0   & 0 & 0\cr
0   & 1 & 0\cr
0   & 0 & 0\cr
0   & 0 & 0\cr
\end{pmatrix},\,
\mathbf{R}_3 =
\begin{pmatrix}
0   & 0 & 0\cr
1   & 0 & 0\cr
0   & 0 & 0\cr
0   & 0 & 1\cr
0   & 0 & 0\cr
\end{pmatrix}
\end{align*}
For example, $\mathbf{R}_1$ corresponds to resource $q = 1$ and describes that it is
(re-)used by link $l = 1$ (first hop of route $i = 1$) and link $l = 5$ (second hop of route $i = 3$).
We will find it useful to define the 2-dimensional equivalent routing matrix,
given by $\mathbf {\tilde{R}} = \sum_{q=1}^Q \mathbf {R}_q$ and entries $\tilde{r}_{li}$.
We assume the data to be routed along a single fixed link, i.e.,
we do not allow the data flow between a Tx-Rx pair to be spread between 2 or more resources.

To describe the association of links with resources, we define the following two functions.
Let $f:I \rightarrow \{1,2\}$ denote the number of hops in the route $i$;
$\mathbf{t}:{I\times \{1,\ldots,f(i)\}} \rightarrow L\times Q$ denote the link and
resource used in route $i$ and hop $h$ respectively. In addition, we denote by $t_1(i,h)$ and $t_2(i,h)$ the
first and second outputs of $\mathbf{t}$, which represent the link and resource respectively.
\TabRef{tab:functions} gives an example of how these functions help to describe the
relationship between routes, links and resource usage.

\begin{table}[!htb]
        \centering
	\caption{An example of how the network in \FigRef{Fig:Model} can be described using the three functions defined above.}\label{tab:functions}
	\footnotesize
	\begin{tabularx}{\columnwidth}{l|X|p{0.27\columnwidth}}
		\hline
		\hline
		\textbf{Function}   & \textbf{Description}                               & \textbf{Example in the Network of Figure \ref{Fig:Model}} \\
		\hline
		\hline
		$f(i)$              & Number of hops in route $i$                        & $f(1) = f(3) = 2$\\ \hline
		$\mathbf{t}(i,h)$   & Link and resource indexes in route $i$ and hop $h$ & $\mathbf{t}(3,2) = (5,1)$\\ \hline
		$t_1(i,h)$          & Link index $l$ in route $i$ and hop $h$            & $t_1(3,2) = 5$\\ \hline
		$t_2(i,h)$          & Resource index $q$ in route $i$ and hop $h$        & $t_2(3,2) = 1$\\
		\hline
		\hline
	\end{tabularx}
\end{table}


\subsection{Assigning a Utility to an S-D Pair}

We let $s_i$ denote the end-to-end {\it rate} for communication
between the S-D pair $i$, which is in correspondence
with the \ac{SINR} {\em targets} for hop $h$ of route $i$ denoted by 
$\gamma_{t_1(i,h)}^{tgt}$.
In a multi-hop communication, the \ac{SINR} targets of each link in a specific route must be the same,
in line with the so-called \emph{solidarity property}~\cite{ChoiAndLee2013}. 
Thus, $\gamma_{t_1(i,h)}^{tgt}$ needs to be indexed with the single index $t_1(i,h)$.

Associated with each S-D pair $i$ is a function $u_i(.)$,
which describes the utility of the S-D pair communicating at rate $s_i$.
We assume that $u_i$ is \textit{increasing} and \textit{strictly concave},
with $u_i\rightarrow -\infty$ as $s_i\rightarrow 0^+$.
In this paper we use $u_i(x) \triangleq \text{ln} (x), \forall i$.

The matrix of link capacities is denoted by $\mathbf {C}=[\mathbf {c}_{1}\cdots \mathbf {c}_q] \in \mathbb{R}^{L\times Q}$,
which depends on the communication bandwidth $W$ of one resource and the {\it achieved} actual \ac{SINR} along route $i$ and
hop $h$, $\gamma_{\mathbf{t}(i,h)}$.
Notice that the achieved \ac{SINR} $\gamma_{\mathbf{t}(i,h)}$ is indexed by $\mathbf{t}(i,h)$, 
because the \acp{SINR} are generally different at different resources.

The vector of total traffic across the links of a route is given by $\mathbf {\tilde{R}}\mathbf {s}$
and the network flow imposes the following set of constraints
on the source-destination rate vector $\mathbf {s}$:
\begin{align*}
\mathbf {\tilde{R}}\mathbf {s} \preceq \sum_{q=1}^Q\mathbf {c}_q\qquad \hspace{1cm}  \mathbf {s} \succeq 0.
\end{align*}
In this formulation,
it is convenient to think of the $\mathbf {s}$ vector as the vector of rates 
while the $\mathbf {c}_q$ vectors represent the Shannon capacity
that can be achieved by the particular power vector $\mathbf {p}_q = [P_{1q},\ldots,P_{Lq}]\in \mathbb{R}^L$ on resource $q$.

Let $G_{\mathbf{t}(i,h)}$ denote the desired link gain on route $i$ and hop $h$,
which includes both large- and small-scale fading gains.
The thermal noise power at the receiver on route $i$ and hop $h$ is denoted by $\sigma_{\mathbf{t}(i,h)}$,
and the transmission power on route $i$ and hop $h$ is $P_{\mathbf{t}(i,h)}$.
The \ac{SINR} on route $i$ and hop $h$ is given by
\begin{equation}
\label{eq:sinr_achiv}
\gamma_{\mathbf{t}(i,h)}(\mathbf{P})=\frac{G_{\mathbf{t}(i,h)}P_{\mathbf{t}(i,h)}}{\sigma_{\mathbf{t}(i,h)} +(P_{\mathbf{t}(i,h)}^{tot}-G_{\mathbf{t}(i,h)}P_{\mathbf{t}(i,h)})}, \nonumber
\end{equation}
where $P_{\mathbf{t}(i,h)}^{tot}$ represent the total received power
measured by the receiver on route $i$ and hop $h$ and $\mathbf {P}=[\mathbf{p}_1,\ldots,\mathbf{p}_Q]\in \mathbb{R}^{L\times Q}$
is the power allocation matrix.

Finally, it will be useful to view each link on route $i$ and hop $h$ as a single Gaussian channel
with Shannon capacity
\begin{equation*}
c_{\mathbf{t}(i,h)}(\mathbf{P}) = W_{t_2(i,h)} \log_2 \left(1 + \gamma_{\mathbf{t}(i,h)}(\mathbf{P})\right),
\end{equation*}
which represents the maximum rate that can be achieved on route $i$
and hop $h$.

\section{Mode Selection and Resource Allocation}\label{sec:ms_ra}

\subsection{Multi-Hop D2D Scenarios: Proximity Communication and Coverage (Range) Extension}
\label{Sec:Scenarios}
Recall from Figure \ref{Fig:Scenario} that \ac{MH} \ac{D2D} communications can be advantageously used
in two distinct scenarios.
In the {\em proximity communication} scenario, a \ac{D2D} relay node helps a \ac{D2D} pair to communicate \cite[Section 5.2.9]{3gpp.22.803},
while in the {\em coverage} or {\em range extension} scenario a \ac{D2D} relay node assists a coverage limited
\ac{D2D} Tx node to boost its link budget to a base station.
In the proximity communication scenario, the mode selection problem consists of deciding whether
the \ac{D2D} Tx node should communicate with the \ac{D2D} Rx node (1) via a direct \ac{D2D} (single-hop) link,
(2) via a 2-hop path through the \ac{D2D} relay node or (3) through the cellular BS.
In contrast, in the range extension scenario, the mode selection problem consists of deciding
whether the \ac{D2D} Tx node should communicate via a direct transmission with its serving BS or
via the \ac{D2D} relay node.
We consider mode selection alternatives in the next subsection.

\subsection{Mode Selection Schemes}
\label{sub:mode_sel}
For the proximity communication scenario,
we use the notion of the equivalent channel from \ac{D2D} Tx to \ac{D2D} Rx through \ac{D2D} relay based on the harmonic mean of the channels from \ac{D2D} Tx to \ac{D2D} relay ($G_{TxRe}$) and from \ac{D2D} relay to \ac{D2D} Rx ($G_{ReRx}$):
\begin{equation}
\label{eq:harmonic}
	\frac{1}{G_{eq}} =  \frac{1}{G_{TxRe}} + \frac{1}{G_{ReRx}}.
\end{equation}
The intuition of defining the equivalent channel according to \eqref{eq:harmonic}
is that the equivalent channel gain tends to be high only when both composite channels are high
and therefore it is an appropriate single measure for mode selection purposes.
A pseudo code of a heuristic mode selection algorithm based on the equivalent
channel is given in \AlgRef{alg:Harmonic}, where we need the channels from the \ac{D2D} Tx to the \ac{BS} ($G_{TxBS}$)
and to the \ac{D2D} Rx ($G_{TxRx}$).
\begin{algorithm}
	\footnotesize
	\caption{\footnotesize \acl{HMS} (HMS) for Proximity Communication}
    \label{alg:Harmonic}
	\begin{algorithmic}[1]
		\IF{$G_{eq} \geq \max\left\{ G_{TxRx},G_{TxBS}\right\}$}
			\STATE Choose \ac{D2D} two-hop communications
		\ELSIF{$G_{TxRx} \geq G_{TxBS}$}
			\STATE Choose \ac{D2D} single-hop communications
		\ELSE
			\STATE Choose cellular mode, that is \ac{D2D} Tx and Rx communication through the BS.
		\ENDIF
	\end{algorithmic}
\end{algorithm}

Recall from Section \ref{Sec:Scenarios} that in the range extension scenario, there are only
two possible communication modes (direct or relay-assisted) between the \ac{D2D} Tx device and the BS.
Therefore, in this scenario, we modify the definition of the equivalent channel such that
it includes the path gain between the relay device and the BS ($G_{ReBS}$):
\begin{equation*}
\label{eq:harmonic2}
	\frac{1}{G_{eq}} =  \frac{1}{G_{TxRe}} + \frac{1}{G_{ReBS}},
\end{equation*}
and use a modified version of the \ac{HMS} algorithm (Algorithm \ref{alg:Harmonic2}).
\begin{algorithm}
	\footnotesize
	\caption{\footnotesize \acl{HMS} (HMS) for Range Extension}
    \label{alg:Harmonic2}
	\begin{algorithmic}[1]
		\IF{$G_{eq} \geq G_{TxBS}$}
			\STATE Choose \ac{D2D} relay assisted communication
		\ELSE
			\STATE Choose cellular mode that is \ac{D2D} Tx transmits directly to the BS.
		\ENDIF
	\end{algorithmic}
\end{algorithm}

\subsection{Resource Allocation Scheme}
\label{sub:res_alloc}
First, we recognize that for two-hop communications with multiple resources the following resource allocation constraints must be met:
\begin{itemize}
 \item A transmitter, either \ac{D2D} Tx ($h=1$) or \ac{D2D} relay ($h=2$), cannot have multiple receivers: $\sum_i \tilde{r}_{t_1(i,h),i} = 1$.
 \item A \ac{D2D} relay cannot receive and transmit on the same resource: $r_{t_1(i,1),i,t_2(i,1)}+ r_{t_1(i,2),i,t_2(i,2)} \leq 1$.
\end{itemize}
Secondly, the set of nodes transmitting to a \ac{BS} must use orthogonal
resources. That is, cellular transmissions maintain intracell orthogonality.
Apart from these constraints,
in this paper we assume that resources are allocated randomly to communication links and leave the study of
efficient resource allocation algorithms for future studies.


\section{Distributed Power Control Optimization}\label{sec:dist_pc}

\subsection{SINR Target Setting and Power Control Problem - Utility Maximization}\label{sub:SINR}
Assuming that the communication-mode has already been selected for the \ac{D2D} candidates,
and all (cellular and \ac{D2D}) links have been assigned a frequency channel or a \ac{RB},
we formulate the problem of target rate setting and power control as:
\begin{equation}
\label{eq:util_max}
\begin{array}{ll}
\displaystyle \maxi_{\mathbf{P},\mathbf{s}} & \sum_i u_i(s_i) - \omega \sum_{i=1}^I\sum_{h=1}^{f(i)} P_{\mathbf{t}(i,h)} \\
\text{subject to} & \mathbf {\tilde{R}}\mathbf {s} \preceq \sum_{q=1}^Q\mathbf {c}_q(\mathbf{P}) ,\quad \forall i,h, \\
& \mathbf{P},\mathbf{s}\succeq 0
\end{array}
\end{equation}
which aims at maximizing the utility while taking into account the transmit powers
(through a predefined weight $\omega \in (0,+\infty) $ \cite{Papandriopoulos2008}),
so as to increase spectrum efficiency while reducing the sum power consumption.

Unfortunately, Problem \eqref{eq:util_max} is not convex. However, exploiting the results presented in \cite{Papandriopoulos2008},
we can transform it into the following equivalent form:
\begin{equation}
\label{eq:util_max_reform}
\begin{array}{ll}
\displaystyle \maxi_{\mathbf{\tilde{s}},\mathbf{\tilde P}}  & \sum_i u_i(e^{\tilde{s}_i}) - \omega \sum_{i=1}^I\sum_{h=1}^{f(i)} e^{\tilde{P}_{\mathbf{t}(i,h)}}  \\
\text{subject to} & \log(\mathbf {\tilde{R}}\mathbf {e^{\tilde{s}}} )\leq \log\left(\sum_{q=1}^Q\mathbf {c}_q(e^\mathbf{\tilde P})\right) \quad \forall i,h,
\end{array}
\end{equation}
where $s_i\leftarrow e^{\tilde{s}_i}$ and $P_{\mathbf{t}(i,h)} \leftarrow e^{\tilde{P}_{\mathbf{t}(i,h)}}$.
The transformed Problem~\eqref{eq:util_max_reform} is proved to be convex
(now in the $\tilde{s}_i$-s and $\tilde{P}_{\mathbf{t}(i,h)}$-s),
for the utility functions $u_i(\cdot)$ are selected to be $(\log,x)$-concave
over their domains \cite{Papandriopoulos2008}.

Under the utility's condition, we can solve Problem~\eqref{eq:util_max_reform}
to optimality by means of decomposing the problem into separate subproblems in $\mathbf{\tilde{s}}$ 
and $\mathbf{\tilde p}$. 
Problem-I~\cite[eq. 5]{Belleschi2009} can be solved by gradient iterations and using Lagrangian duality to obtain the \ac{SINR} targets,
while Problem-II~\cite[eq. 8]{Belleschi2009} can be solved by an iterative \ac{SINR} target following inner loop
(set by a Zander type iterative \ac{SINR} target~\cite{Zander1992}).
The relationship between Problem-I and Problem-II can be exploited
such that the necessary Lagrange multipliers in the iterations of Problem-I
are provided by solving Problem-II. The details are omitted here due to space limitations.


\section{Numerical Results and Discussion}
\label{sec:num_res}


\subsection{Simulation Setup and Parameters}
\label{sub:sim_set}
In this section, we consider a seven cell system with a cell radius of 500 m supporting
18 uplink physical \acp{RB} in each cell.
The \ac{D2D} communication uses uplink \acp{RB} in both the proximity communication
and the range extension scenarios.
For simplicity and to gain insights, we assume that each \ac{UE} and \ac{D2D} pair
uses a single uplink \ac{RB}.
The most important system parameters are summarized in Table \ref{tab_sim_param}.

\begin{table}[t]
	\caption{Simulation parameters}
    \label{tab_sim_param}
	\footnotesize
	\begin{tabularx}{\columnwidth}{X|X}
		\hline
		\hline
		\textbf{Parameter}                     & \textbf{Value} \\
		\hline
		\hline
		Number of \acp{BS}                    & 7  \\
		Cell radius			                   & \unit{500}{m} \\
		Minimum distance \ac{BS}-\ac{UE}      & \unit{50}{m} (Scen. 1)/\unit{400}{m} (Scen. 2)\\
		Minimum distance \ac{UE}-\ac{UE}       & \unit{10}{m} \\
		Mean distance \ac{D2D} Tx-\ac{D2D} Rx          & \unit{100}{m} \\
		Number of cellular \acp{UE} per cell   & 6  \\
		Number of \ac{D2D} triplets per cell   & 6 (Scen. 1)/18 (Scen. 2)  \\
		Monte Carlo iterations                 & $100$ \\ \hline		
		Central carrier frequency              & \unit{2}{GHz} \\
		System bandwidth					   & \unit{5}{MHz} \\
		Number of \acp{RB}                     & 18 \acp{RB}  \\
		Gain at \unit{1}{m} distance		   & \unit{-37}{dB} \\
		Thermal Noise power per \ac{RB}		   & \unit{-116.4}{dBm} \\
		Path Loss coefficient				   & 3.5 \\
		Shadowing standard deviation           & \unit{8}{dB} \\ \hline
		\ac{BS} transmit power                & \unit{40}{dBm} \\
		\ac{UE} min/max transmit power 		   & \unit{-23}{dBm}/\unit{23}{dBm}\\
		Fixed Power for \ac{LTE} \ac{PC}     & \unit{-10}{dBm}\\
		Path loss compensation factor ($\alpha$)& 0.8\\
		\ac{SNR}/\ac{SINR} target	  		   & \unit{15}{dB} \\
		Number of outer-loop iterations		   & 70 \\
		Number of inner-loop iterations		   & 10 \\
		$\epsilon$ for the outer-loop 		   & 0.05 \\
		Initial power for the inner-loop       & \unit{10}{dBm} \\
		Initial $\gamma^{tgt}$ for the outer-loop& \unit{0}{dB} \\
		$\omega$ of Eq. \eqref{eq:util_max}		 & [0.1 1 10 100] \\
		\hline
		\hline
	\end{tabularx}
\end{table}

To collect statistics on the measured \ac{SINR} and transmit power levels, we perform
Monte Carlo simulations, such that in each Monte Carlo experiment we randomly drop
6 {\it cellular} \acp{UE} and 6 \ac{D2D} {\it triplets} per cell for the proximity communication
scenario and 18 \ac{D2D} {\it triplets} per cell for the range extension scenario.
A cellular \ac{UE} refers to a \ac{UE} that transmits to its serving \ac{BS}, while a triplet is
a set consisting of a \ac{D2D} transmitter, a \ac{D2D} relay and a \ac{D2D} receiver node.
Recall that in the proximity communication scenario a \ac{D2D} transmitter
transmits to a \ac{D2D} receiver node (possibly via a \ac{D2D} relay), while in the
range extension scenario, a \ac{D2D} transmitter node transmits to its serving \ac{BS}
(possibly via a \ac{D2D} relay).
In the range extension scenario, the \ac{D2D} receiver node is not used.

\begin{table}[!ht]
    \centering
	\caption{Mode selection algorithms}
    \label{tab:ms}
	\footnotesize
	\begin{tabularx}{\columnwidth}{X|X|X}
		\hline
		\hline
		\textbf{Name}                  & \textbf{Proximity Communications Scenario}             & \textbf{Range Extension Scenario} \\
		\hline
		\hline
		Cellular mode (\textbf{Cmode})          & Forced cellular mode (no \ac{D2D} communications)   & Forced cellular mode (no \ac{D2D} communications) \\ \hline
		\ac{D2D} mode (\textbf{DMS})            & Mode selection between single-hop \ac{D2D} mode and cellular mode     & Forced relaying (two hop) \ac{D2D} mode, that is transmission through the \ac{D2D} relay node \\ \hline
        Adaptive mode selection with the HMS algorithms (\textbf{\ac{HMS}})  & Mode selection by \textbf{Algorithm 1}         & Mode selection by \textbf{Algorithm 2} \\
		\hline
		\hline
	\end{tabularx}
\end{table}

To gain insight into the performance impacts of mode selection algorithms,
we evaluate the mode selection (MS) alternatives listed in Table \ref{tab:ms}.

\begin{table}[!ht]
    \centering
	\caption{Power control algorithms}
    \label{tab:pc}
	\footnotesize
	\begin{tabularx}{\columnwidth}{l|X|X}
		\hline
		\hline
		\textbf{Name}                  & \textbf{Cellular UE power control}                     & \textbf{\ac{D2D} power control} \\
		\hline
		\hline
		\textbf{Fix}                            & LTE Open Loop         & Fixed Power \\ \hline
		\textbf{Fix SNR}                        & LTE Open Loop         & Fixed \ac{SNR} target \\ \hline
		Open Loop \textbf{(OL)}                 & LTE Open Loop         & LTE Open Loop \\ \hline
		Closed Loop \textbf{(CL)}               & LTE Open Loop         & LTE Closed Loop \\ \hline
    Utility Maxim. \textbf{(UM-$\omega$)}       & Utility maximizing PC with parameter $\omega$         & Utility maximizing PC with parameter $\omega$ \\
		\hline
		\hline
	\end{tabularx}
\end{table}

To evaluate and benchmark the performance of the utility maximizing power
control scheme, we compare its \ac{SINR} and power consumption statistics with those
based on the well known \ac{LTE} power control schemes~\cite{R1-074850}, 
as listed in Table \ref{tab:pc}.

\subsection{Impact of Mode Selection Algorithms}
\label{sub:perf_analy_ms}
\begin{figure}[ht!]
\centering
\includegraphics[width=0.85\hsize]{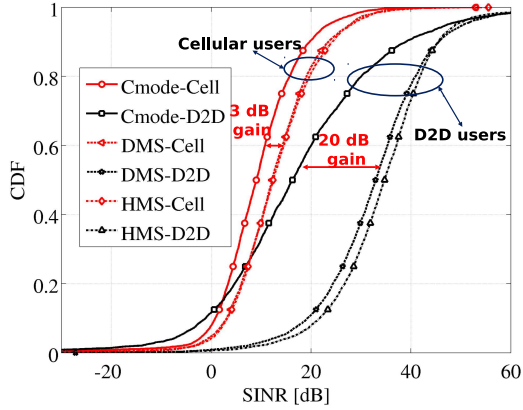}
\caption{\textbf{Proximity communication scenario:}
\ac{CDF} of the \ac{SINR} for both cellular \acp{UE} and \ac{D2D} candidates with Cmode, \acs{DMS} and \ac{HMS}
(see \TabRef{tab:ms}).
\ac{HMS} is superior for both the cellular \acp{UE} (denoted '-Cell') and the \ac{D2D} candidates and considering all the modes.
The cellular \acp{UE} benefit somewhat ($\approx$ \unit{3}{dB}) from \ac{D2D} communications.
For the \ac{D2D} candidates, the mode selection gain is much more pronounced ($\approx$ \unit{22}{dB}) with the \ac{HMS}.
}
\label{fig:2hop_d2d_joint_users}
\end{figure}

\begin{figure}[ht!]
\centering
\includegraphics[width=0.80\hsize]{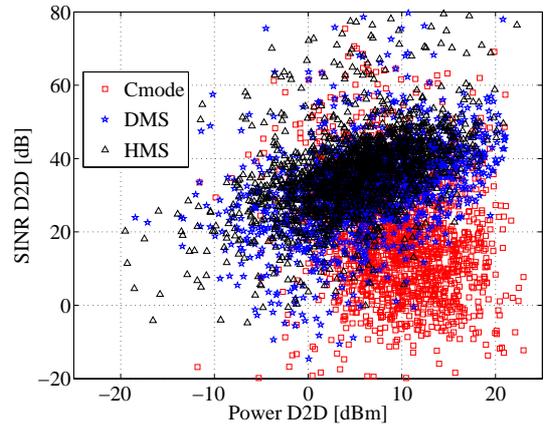}
\caption{\textbf{Proximity communication scenario:}
\ac{CDF} of the \ac{SINR} for both cellular \acp{UE} and \ac{D2D} candidates when considering different communication modes.
We notice that Cmode results in lower \ac{SINR} values with a higher power consumption
than all the other modes. In addition, \ac{HMS} reaches higher \ac{SINR} values than single hop \ac{D2D} mode
with a lower power consumption, which suggests that in addition to the \ac{SINR} gains,
two-hop communications outperform the single-hop \ac{D2D} mode.}
\label{fig:2hop_d2d_scatter_d2d}
\end{figure}

\begin{figure}[ht!]
\centering
\includegraphics[width=0.85\hsize]{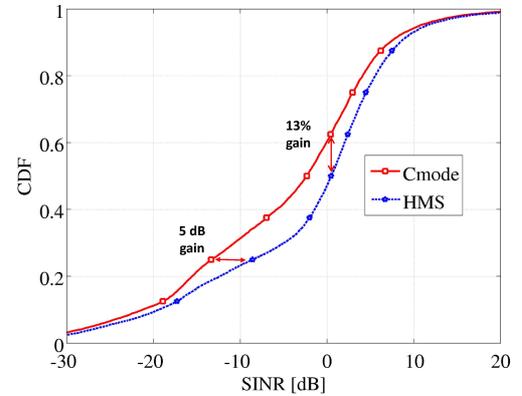}
\caption{\textbf{Range extension scenario:}
\ac{CDF} of the \ac{SINR} for \ac{D2D} candidates when considering different communication modes. We notice
that the \ac{HMS} outperforms the Cmode in the low \ac{SINR} regime. Moreover, \ac{HMS}
decreases the occurrence of \ac{SINR} values below \unit{0}{dB}.}
\label{fig:2hop_ce_joint_users}
\end{figure}

Figures \ref{fig:2hop_d2d_joint_users}-\ref{fig:2hop_ce_joint_users} compare the
performance of the forced cellular mode,
\ac{D2D} mode (mode selection between single hop and cellular communications)
and \ac{HMS} (see Table \ref{tab:ms}).

\FigRef{fig:2hop_d2d_joint_users} shows the \ac{SINR} distributions of cellular \acp{UE} and
\ac{D2D} pairs when employing the mode selection schemes of Table \ref{tab:ms}
in the proximity communication scenario.
This figure shows that cellular \acp{UE} (transmitting to their serving BS) benefit somewhat ($\approx$ \unit{3}{dB})
from \ac{D2D} communications, especially when adaptive mode selection (the \ac{HMS} algorithm) is used for mode selection.
For the \ac{D2D} users the mode selection gain is much more pronounced ($\approx$ \unit{20}{dB}).
The intuitive explanation of this is that D2D communication with adaptive power control takes advantage of the proximity
gain and reduces intercell interference.
At the same time, \ac{D2D} \acp{UE} benefit from an improved  link budget due to the proximity,
which allows for lower transmit power and higher \ac{SINR} at the D2D receivers.
\ac{HMS} can adaptively take advantage of the two-hop path,
which explains the additional gain of HMS over DMS ($\approx$ \unit{2}{dB}).

Figure \ref{fig:2hop_d2d_scatter_d2d} is the scatter plot
of the transmit power levels and achieved \ac{SINR} levels of \ac{D2D} candidates
in the proximity communication scenario, which shows that Cmode results
in lower \ac{SINR} values with a higher power consumption than all the other modes.
Also, \ac{HMS} reaches higher \ac{SINR} values than single-hop \ac{D2D} mode with a lower power consumption,
which suggests that in addition to the \ac{SINR} gains,
two-hop communications outperform single-hop \ac{D2D} mode in terms of power efficiency.


\FigRef{fig:2hop_ce_joint_users} shows the \ac{SINR}
distribution for the \ac{D2D} nodes using the Cmode and the \ac{HMS}
mode selection algorithm in the range extension scenario.
\FigRef{fig:2hop_ce_joint_users} shows that the \ac{HMS} outperforms Cmode with margin of \unit{5}{dB} in the low \ac{SINR} regime.
The \acp{UE} that experience low \ac{SINR} values
are the ones at the cell edge
and benefit the most from the presence of \ac{D2D} relay nodes.
In addition, \ac{HMS} reduces the probability of the \ac{SINR} being below \unit{0}{dB} from \unit{60}{\%} to \unit{47}{\%}.
This is because the mode selection algorithm exploits the fact that the \ac{MH} path is stronger than in the Cmode,
and thus yields a proximity gain for cell edge users.

Our conclusion regarding mode selection algorithms is that
both proximity communication and \ac{D2D} range extension
can benefit from MH communication in terms of spectral and energy efficiency
when the communication mode is properly (that is adaptively) selected.

\subsection{Impact of Power Control Algorithms}
\label{sub:perf_analy_pc}
To gain insight into the impact of power control,
we consider the power control algorithms of Table \ref{tab:pc} using \ac{HMS} for both
the proximity communication and range extension scenarios.
For the utility maximization power control scheme, we employ
four different values of $\omega$, which controls the spectral and energy efficiency trade-off.

\begin{figure}[ht!]
\centering
\includegraphics[width=0.85\hsize]{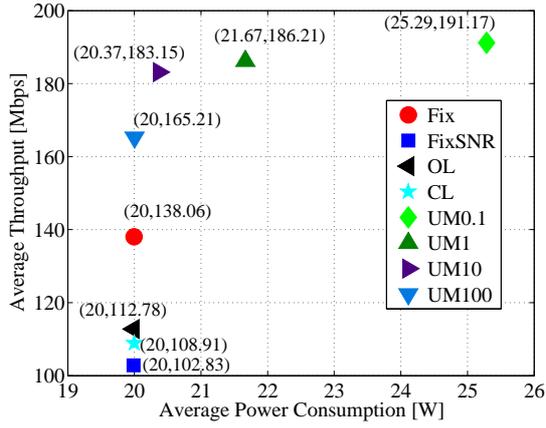}
\caption{\textbf{Proximity communication scenario:}
Scatter plot of the total power consumption and average throughput achieved by the examined power control algorithms. $(x,y)$ near each symbol shows the $x$-axis (power consumption in W) and $y$-axis (throughput in Mbps) values.
Note that UM10 can boost the average throughput with a small increase of the transmit power level.
}
\label{fig:2hop_d2d_scatter_rate_power}
\end{figure}

\FigRef{fig:2hop_d2d_scatter_rate_power} is the scatter plot for the proximity communication scenario.
With $\omega=0.1$ the average throughput gain is approximately \unit{39}{\%} over the \ac{LTE} \ac{PC}
with fixed power, but using approximately \unit{26}{\%} more power.
However, with $\omega=100$ the average throughput gain is approximately \unit{20}{\%}
using similar transmit power levels as LTE PC.
It is interesting to note that the utility maximizing \ac{PC} can trade between different objectives while maintaining high
values of average throughput compared with all the other \ac{PC} algorithms.

\begin{figure}[ht!]
\centering
\includegraphics[width=0.85\hsize]{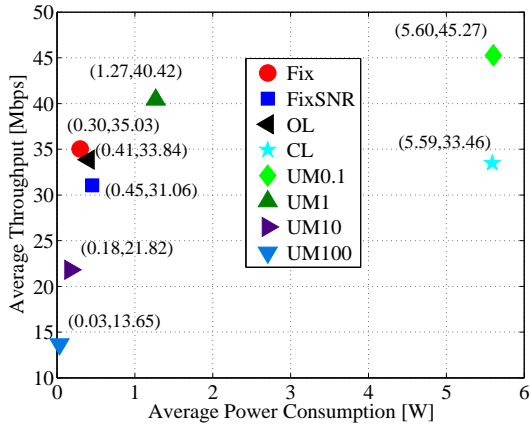}
\caption{\textbf{Range extension scenario:}
Scatter plot of the total power consumption and average throughput achieved by the examined power control algorithms.
The utility maximizing \ac{PC} can reach the highest throughput (with lower values of $\omega$) or
the lowest power consumption (with higher $\omega$ values). LTE OL provides a reasonable engineering trade-off.}
\label{fig:2hop_ce_scatter_rate_power}
\end{figure}

\FigRef{fig:2hop_ce_scatter_rate_power} is the scatter plot for the range extension scenario.
Similarly to the previous figure, for $\omega=0.1$ the utility maximization reaches the highest average throughput,
with a gain of approximately \unit{29}{\%} over \ac{LTE} \ac{PC} with fixed power.
However, with $\omega\geq 10$ the utility maximizing PC minimizes
power consumption
at the expense of reaching lower throughput values.
Clearly, utility maximizing \ac{PC} can reach high throughput when using low values of $\omega$.
However, if the power consumption has to be kept at low values with reasonable throughput values,
utility maximization with higher $\omega$ values or using the \ac{LTE} \ac{PC} can be satisfactory.
\vspace{-2mm}
\section{Conclusion}\label{sec:concl}
In this paper we developed radio resource management algorithms applicable in network-assisted
\ac{MH} \ac{D2D} scenarios, including the proximity communication and the range extension scenarios.
The proposed adaptive harmonic mode selection (HMS) scheme together with a utility maximizing
PC scheme can improve
the throughput and the energy efficiency of a system that does not support
D2D communications or employs traditional mode selection and power control schemes.
HMS can also decrease the outage probability and improve the average throughput
using similar transmit power levels as users employing traditional PC techniques.
LTE OL power control can also provide a reasonable trade-off between throughput
and energy efficiency, especially in the range extension MH scenario.
\vspace{-2mm}
\bibliographystyle{IEEEtran}
\bibliography{references,IEEEfull}


\end{document}